\documentclass[aps,prl,twocolumn,floatfix]{revtex4-2}
\usepackage[utf8]{inputenc}
\usepackage[english]{babel}
\usepackage{mathtools}
\usepackage{newtxtext}  % scaled=1.005
\usepackage{amsmath}
\usepackage{amsfonts}
\usepackage[varbb,varg]{newtxmath} % scaled=1.005
\usepackage{bm,dsfont,eucal}
\AtBeginDocument{\renewcommand{\vec}[1]{\bm{#1}}}

\usepackage[dvipsnames,svgnames,x11names,hyperref]{xcolor}
\usepackage{hyperref}
\hypersetup{ % closer to what APS journals use
    colorlinks=true, linkcolor=NavyBlue, urlcolor=NavyBlue, citecolor=NavyBlue
}
\usepackage{physics,braket,siunitx,slashed}
\usepackage[capitalise]{cleveref}

\begin{document}
        \title{Unconventional Metallic Ferromagnetism:
        Non-Analyticity and Sign-Changing Behavior of Orbital Magnetization in Rhombohedral Trilayer Graphene
%Non-Analyticity and Sign-Changing Behavior of Magnetization in an Orbital Chern Metal
%Revealing the Non-Analytic Nature and Sign-changing Landscape of Metallic Orbital Magnetization: A case study with quarter metal ABC Trilayer Graphene
} 

\author{Mainak Das and Chunli Huang}
\affiliation{Department of Physics and Astronomy, University of Kentucky, Lexington, Kentucky 40506-0055, USA}

\date{\today} 

\begin{abstract}
We study an unique form of metallic ferromagnetism in which orbital moments surpasses the role of spin moments in shaping the overall magnetization.  This system emerges naturally upon doping a topologically non-trivial Chern band in the recently identified quarter metal phase of rhombohedral trilayer graphene.
Our comprehensive scan of the density-interlayer potential parameter space reveals an unexpected landscape of orbital magnetization marked by two sign changes and a line of singularities.  The sign change originates from an intense Berry curvature concentrated close to the band-edge, and the singularity arises from a topological Lifshitz transition that transform a simply connected Fermi sea into an annular Fermi sea.
Importantly, these variations occur while the groundstate order-parameter (i.e.~valley and spin polarization) remains unchanged. 
This unconventional relationship between the order parameter and magnetization markedly contrasts traditional spin ferromagnets, where spin magnetization is simply proportional to the groundstate spin polarization via the gyromagnetic ratio. 
We compute energy and magnetization curves as functions of collective valley rotation to shed light on magnetization dynamics and to expand the Stoner-Wohlfarth magnetization reversal model. We provide predictions on the magnetic coercive field that can be readily tested in experiments.
Our results challenge established perceptions of magnetism, emphasising the important role of orbital moments in two-dimensional materials such as graphene and transition metal dichalcogenides, and in turn, expand our understanding and potential manipulation of magnetic behaviors in these systems.
\end{abstract} 

\maketitle

\textit{Introduction: }Ferromagnetic and antiferromagnetic behaviors of most materials are principally guided by the Pauli exclusion principle, which influences the momentum and spatial distributions of electrons based on their relative spin orientations \cite{herring1966magnetism}. Traditionally, magnetization in these materials has been attributed to the accumulation of spin moments, while the contribution of orbital moments induced by spin-orbit coupling has often been regarded as secondary or negligible. However, this conventional understanding changes in two-dimensional materials like graphene, where electrons in the low-energy regime possess an additional degree of freedom -- known as valley -- that is intrinsically associated with orbital moments. 
The interplay between the Pauli exclusion principle and the expanded four-dimensional spin-valley space in graphene gives rise to an unconventional form of magnetism, wherein orbital moments can compete or even surpass the role of spin moments in determining the overall magnetization.

%The understanding of magnetism in most materials has long been centered around the role of spin moments governed by the Pauli exclusion principle. Orbital moments induced by spin-orbit coupling were typically considered secondary or negligible in this context. However, the landscape changes in two-dimensional materials like graphene, where electrons in the low-energy regime possess an additional degree of freedom called valley, associated with orbital moments. The interplay between spin and valley degrees of freedom within the expanded 4-dimensional spin-valley space of graphene gives rise to a distinct form of magnetism where orbital moments can rival or even surpass the significance of spin moments.

Experimental studies conducted in recent years have indeed demonstrated that manipulating multilayer graphene with electric displacement fields  \cite{zhou_half_2021,zhou_isospin_2021,seiler2022quantum,han2023correlated} or moire potentials \cite{he2021competing,chen2021electrically,tschirhart2021imaging,sinha2022berry,pantaleon2022superconductivity} can lead to both metallic and insulating magnetism, where orbital moments play a dominant or equally important role as spin moments in determining the overall magnetization. While significant research efforts have focused on insulating orbital magnets \cite{polshyn2020electrical,polshyn2020electrical,wang2021phase,huang2021current,zhu2020voltage,ren2021orbital,zhao2022realization,liu2021orbital,tao2022valley,qiu2023interaction}, this study shifts the focus to exploring the behavior of orbital magnetization in the metallic regime. Specifically, we study the quarter metal phase of rhombohedral trilayer graphene within the experimentally accessible density-interlayer potential ($n_e-U$) parameter space. 
This system has the advantage of an experimentally tunable balance between Coulomb and band energies. %Moreover, it represents a simple realization of an itinerant electron orbital magnet from doping a topologically non-trivial Chern band.  
\begin{figure}[t]
    \centering
    \includegraphics[width=\columnwidth]{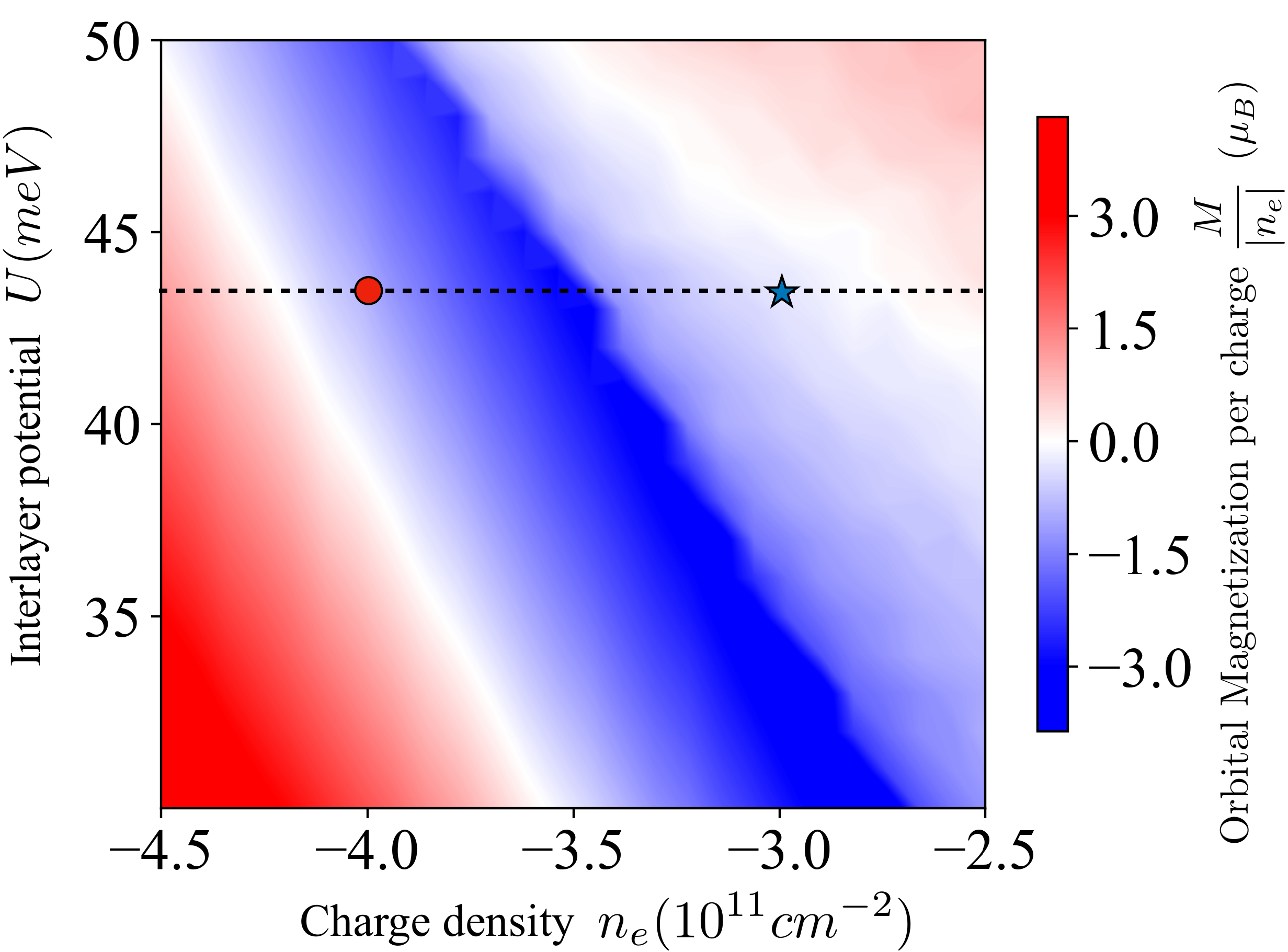}
    \caption{
    The orbital magnetization, $M$, of the quarter metal displays a rich landscape across the $n_e-U$ parameter space, even as the ground state order parameter (i.e. valley polarization) remains unchanged. In the white region where $M$ vanishes, the groundstate is unresponsive to weak orbital magnetic field. Along the blue diagonal line, where $M$ reaches a minimum, it exhibits non-analytic behavior due to a topological Lifshitz transition. Fig.~\ref{fig:2} further illustrates $M$ v.s.~$n_e$ along the $U=43$meV horizontal dotted line and the bandstructure at two specific points.
    }
    \label{fig:1}
\end{figure}

\begin{figure*}[t]
    \centering
    \includegraphics[width=2.05\columnwidth]{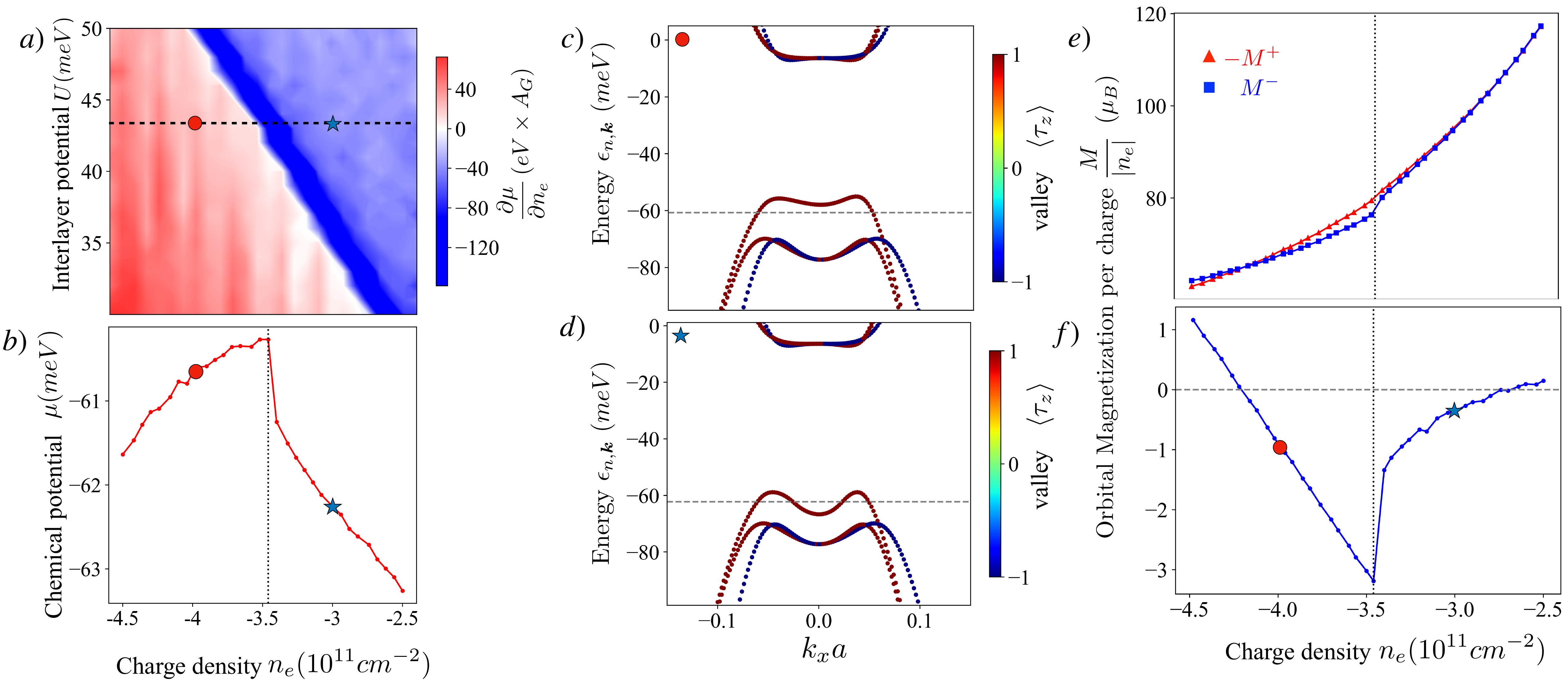}
    \caption{$\vec{a)}$ Inverse compressibility $\partial \mu /\partial n_e$ v.s. $n_e$ exhibits a line of singularities where annular Lifshitz transition (ALT) happens. $\vec{b)}$ $\mu$ vs $n_e$ along the $U=43~meV$ shows a singularity at $n_{ALT}=-3.4\times 10^{11} cm^{-2}$. $\vec{c},\vec{d)}$ The electronic bandstructure of valley-Ising metal with a simply connected Fermi sea  and annular Fermi sea. The horizontal dashed line represent the corresponding chemical potential$(\mu)$. $\vec{e)}$ and $\vec{f)}$: Majority valley ($M^{-}$), minority valley ($M^+$) and total $M=M^+ + M^-$ orbital magnetization per charge carrier vs $n_e$ at $U=43 ~meV$. 
    % Note  $d(M/n_e)/dn_e$ shows a more pronounced discontinuity at the ALT compare to $d\mu/dn_e$.
    % \textcolor{red}{Suggestion. Change y-axis to  $M/|n_e|$ ($mu_B$)}
    %Chemical potential and orbital magnetization along the horizontal dashed line $(U=43 meV)$ shown in Fig 1 and 2.a,b. 
    }
    \label{fig:2}
\end{figure*}

In this Letter, we uncover a diverse landscape of orbital magnetization $M$ in the $n_e-U$ space of the quarter metal in rhombohedral trilayer graphene, featuring two sign changes and a line of discontinuity, as depicted in Fig.~\ref{fig:1}. Importantly, these variations occur while the groundstate order parameter (valley and spin polarization) remains unchanged.
This independent relation between the order parameter and magnetization starkly deviates from conventional ``spin'' ferromagnets, where magnetization arises primarily from the alignment of spin moments and thus is directly proportional to the groundstate spin polarization via the gyromagnetic ratio.
%The discontinuity occurs when a simply connected Fermi surface undergoes a topological Lifshitz transition, leading to the formation of an annular Fermi sea.  
In addition to exploring the experimentally relevant $n_e-U$ space, we compute the energy and magnetization curve as a function of collective valley rotation to provide insights into the dynamical behavior of orbital magnetism. From this, we compute magnetic coercive field which can be readily verified by future experiments.

%Overall, our findings shed light on the unique properties of orbital magnetism in quarter metal rhombohedral trilayer graphene, showcasing its potential for electric control and manipulation.

\textit{Valley-Ising Metal and Orbital Magnetization:}
We use self-consistent mean-field theory 
\cite{huang2023spin,chatterjee2022inter} to study orbital magnetization in the quarter metal phase of ABC trilayer graphene. For a given point in the $n_e-U$ space, we solve the eigenvalue equation $\hat{H}_{\vec{k}}|\psi_{n\vec{k}}\rangle=\epsilon_{n\vec{k}}|\psi_{n\vec{k}}\rangle$where the mean-field Hamiltonian
\begin{equation} \label{eq:H_MF}
    \hat{H}_{\vec{k}} = \hat{T}_{\vec{k}} + \hat{\Sigma}^{F}_{\vec{k}},
\end{equation}
%$V_{\bold{q}}=2\pi k_e \tanh(|\bold{q}|d)/(\epsilon_r |\bold{q}|)$ from  arXiv:2203.12723v1 (Mainak)
accounts for the experimentally informed \cite{zhou_half_2021} Slonczewski-Weiss-McClure parameters in the band-Hamiltonian $\hat{T}_{\vec{k}}$ \cite{supmat} and the influence of gate-screened Coulomb interaction with Fourier components $V_{\bold{q}}=2\pi k_e \tanh(|\bold{q}|d)/(\epsilon_r |\bold{q}|)$ in the Fock self-energy $\hat{\Sigma}^{F}_{\vec{k}}=-\frac{1}{\mathcal{A}}\sum_{n\vec{k}'}V_{\vec{k}-\vec{k}'} n_{F}(\epsilon_{n\vec{k}'})|\psi_{n\vec{k}'}\rangle\langle \psi_{n\vec{k}'}|$. Note the interlayer potential $U$ is a parameter in $\hat{T}_{\vec{k}}$. $\mathcal{A}$ is area of the sample, $k_e=1.44$eVnm is the Coulomb constant, $\epsilon_r$ is the screening constant, $d$ is the distance from the gate to the material and $n_{F}(\epsilon_{n\vec{k}})=1/(1+\exp(\beta (\epsilon_n(\vec{k})-\mu)))$ is Fermi-Dirac distribution function. At each $\vec{k}$ point, $\hat{H}_{\vec{k}}$ is a $24\times24$ matrix, accounting for the $6$ sublattices, $2$ spin (s), and $2$ valley ($\tau$) degrees of freedom. A high-resolution $k$-mesh of $160\times160$ is used to capture the details of the Fermi surface and provide energy resolution between distinct competing states. 

In the $n_e-U$ phase space presented in Fig.~\ref{fig:1}, with $\epsilon_r=15$ and $d=5$ nm, we found the groundstate is a fully spin-polarized quarter metal. The associated valley-polarization, $\tau_z\equiv \cos\theta_v$, is either $\theta_v=0$ or $\theta_v=\pi$, leading us to term the groundstate a valley-Ising quarter metal. This state undergoes an Annular Lifshitz Transition (ALT) which has a simply connected Fermi sea on the high-density side (Fig.~\ref{fig:2}c) and an annular Fermi sea on the low-density side (Fig.~\ref{fig:2}d).
As shown in Fig.~\ref{fig:2}a, the mean-field theory agrees qualitatively with the experimental observations of the ALT phase boundary in the $n_e-U$ space, as well as the order of magnitude for the inverse compressibility $\partial \mu/\partial n_e$ \cite{zhou_half_2021}.
%The negative compressibility observed in the low-density side of the ALT is due to the substantial enhancement of the exchange energy generated by the newly created inner Fermi surface. 
The negative compressibility observed on the low-density side of the ALT stems from exchange energy enhancement, triggered by the formation of a new inner Fermi surface.
A negative compressibility was observed in two-dimensional electron gas when exchange energy dominates the kinetic energy at low density ~\cite{eisenstein1994compressibility}.
%, cite Einsenstein paper.

While we do not anticipate mean-field theory to yield accurate results across the entire $n_e-U$ phase diagram \footnote{In preparation. Tobias Wolf, Chunli Huang, Allan MacDonald}, this agreement suggests that mean-field theory captures the key features of the quarter metal phase.
These features are attributed to the simplicity of the Fermi surface coupled with the favorable electron density and strong Coulomb repulsion, enabling the distribution of holes into a single spin-valley flavor. Indeed, the primary spectral density of the metallic band (with Fermi energy $\sim 5$meV) has a large energy separation from above driven mainly by the interlayer potential ($\sim40$meV) and from below driven by the exchange energy ($\sim20$meV).

Having elucidated the bandstructure of the valley-Ising metal, we now shift our focus to its magnetic properties by using the so-called modern theory of orbital magnetization \cite{hanke2016role,shi2007quantum,thonhauser2011theory,xiao2021adiabatically,PhysRevLett.95.137205,PhysRevLett.110.087202,ceresoli2010first,ceresoli2006orbital,resta2010electrical,raoux2015orbital,piechon2016geometric,PhysRevB.102.184404,PhysRevB.100.054408,PhysRevB.105.195426}. 
%\textcolor{red}{Insert all relevant CITATION,we dont want to upset any party}
Following Ref.~\cite{shi2007quantum}, we derive the formula for $M$ in terms of the mean-field eigenstates and eigenvalues by computing the expectation value of the Hamiltonian at linear order in the minimal-coupling $V_B=e\sum_{i=x,y}\{\hat{v}_i,A^i\}/2$ using standard perturbation theory. Here, $\hat{v}_i=\partial_{k_i} \hat{H}$ is the velocity operator and $A^i$ is the gauge field. Specifically, we equate the magnetization energy, $-MB$, to the first-order energy change  $\delta E =\text{Tr}[\delta \rho \hat{H}]$ where $\delta \rho$ is the first-order change in the density matrix. Importantly, it is the change of the Bloch wavefunction $|\delta\psi_{n\vec{k}}\rangle$, as opposed to changes in the distribution function, that contributes at this order,
$\delta \rho =\sum_{n\vec{k}} n_{F}(\epsilon_{n\vec{k}})\left( |\delta\psi_{n\vec{k}}\rangle\langle \psi_{n\vec{k}}|+\text{h.c.}\right)
$ where
\begin{equation} \label{eq:del_psi}
    |\delta \psi_{n,\vec{k}}\rangle = \sum_{n'\neq n,\vec{k}\neq\vec{k'}}\frac{\langle\psi_{n',\vec{k'}}|\hat{V}_B|\psi_{n,\vec{k}}\rangle}{\left(\epsilon_{n,\vec{k}}-\epsilon_{n',\vec{k'}}\right)} |\psi_{n',\vec{k'}}\rangle.
\end{equation}
As a result, $M$ is sensitive not only to the properties at the Fermi surface but also to the entire eigenvalue spectrum. We refer the readers to  Ref.~\cite{supmat,shi2007quantum} for more discussion and present the final expression for zero-temperature valley-projected orbital magnetization (per area) $M^{\tau}$:
%\begin{equation} \label{eq:OM}
%      = \frac{e}{\hbar}\operatorname{Im}\sum_{n,\vec{k},n'\neq n}f_{n,\vec{k}} \frac{\langle {n,\vec{k}}|\frac{\partial \hat{H}}{\partial \vec{k}}|{n',\vec{k}}\rangle & \times \langle {n',\vec{k}}|\frac{\partial \hat{H} }{\partial\vec{k}}|{n,\vec{k}}\rangle}{(\epsilon_{n,\vec{k}}-\epsilon_{n',\vec{k}})^2}\\  
%      &(\epsilon_{n,\vec{k}}+\epsilon_{n',\vec{k}}-2\mu)\
% \end{equation}

% \begin{align} \label{eq:OM}
%  M_\tau =
%  &- \frac{\mu}{|n_e|}  \sum_{n}^{'} \int \frac{d^2k}{(2\pi)^2} \bigg[
%  n_F(\epsilon_{n\vec{k}}^{\tau}) \Omega_{n\vec{k}}^{\tau}   \bigg] \nonumber \\
%  &+ \frac{e }{2 \hbar |n_e|}\sum_{n}^{'}
%   \int \frac{d^2k}{(2\pi)^2}   \bigg[
%   n_{F}(\epsilon_{n\vec{k}}^{\tau})(\epsilon_{n,\vec{k}}^{\tau}+\epsilon_{n',\vec{k}}^{\tau})  \nonumber \\
%   \; & \times \sum_{n'\neq n}
%  \frac{ \epsilon_{ij} \operatorname{Im} \left( \langle \psi_{n,\vec{k}}^{\tau}| v_i |\psi_{n',\vec{k}}^{\tau}\rangle  \langle \psi_{n',\vec{k}}^{\tau} |v_j| \psi_{n,\vec{k}}^{\tau} \rangle \right)}{(\epsilon_{n,\vec{k}}^{\tau}-\epsilon_{n',\vec{k}}^{\tau})^2} \bigg] 
% \end{align}
\begin{align} \label{eq:OM}
 M^\tau =& \; \frac{e\mu}{\hbar}  \sum_{n=1}^{12} \int \frac{d^2k}{(2\pi)^2} \bigg[
n_F(\epsilon_{n\vec{k}}^{\tau})\Omega_{n\vec{k}}^{\tau}  \bigg] \nonumber \\
 +& \frac{e }{2 \hbar  } 
 \sum_{\substack{
   n=1}}^{12}
  \int \frac{d^2k}{(2\pi)^2}   \sum_{\substack{n'=1 \\ n' \neq n}}^{12} \bigg[
  n_{F}(\epsilon_{n\vec{k}}^{\tau})(\epsilon_{n,\vec{k}}^{\tau}+\epsilon_{n',\vec{k}}^{\tau})  \nonumber \\
  \; & \times 
 \frac{ \epsilon_{ij} \operatorname{Im} \left( \langle \psi_{n,\vec{k}}^{\tau}| \hat{v}_i |\psi_{n',\vec{k}}^{\tau}\rangle  \langle \psi_{n',\vec{k}}^{\tau} |\hat{v}_j| \psi_{n,\vec{k}}^{\tau} \rangle \right)}{(\epsilon_{n,\vec{k}}^{\tau}-\epsilon_{n',\vec{k}}^{\tau})^2} \bigg].
\end{align}
Here $\epsilon_{ij}=-\epsilon_{ji}$ is the antisymmetric tensor and $\Omega^\tau$ is the valley-projected Berry-curvature
\begin{equation}
    \Omega_{n\vec{k}}^{\tau}=-\sum_{\substack{
   n'=1 \\
   n' \neq n
  }}^{12}  
 \frac{ \epsilon_{ij} \operatorname{Im} \left( \langle \psi_{n,\vec{k}}^{\tau}| \hat{v}_i |\psi_{n',\vec{k}}^{\tau}\rangle  \langle \psi_{n',\vec{k}}^{\tau} |\hat{v}_j| \psi_{n,\vec{k}}^{\tau} \rangle \right)}{(\epsilon_{n,\vec{k}}^{\tau}-\epsilon_{n',\vec{k}}^{\tau})^2},
\end{equation}
which integrates to $2\pi$ times the Chern number $C^{\tau } = \pm 3$ when the valley-projected valence bands are completely occupied.

Fig.~\ref{fig:2}e shows the magnetization curves as a function of $n_e$ for the majority-valley $M^{-}>0$ and minority-valley $M^{+}<0$.  Both curve exhibits a strong dependence on $n_e$, largely dictated by the contribution of Berry curvature (first term in Eq.\eqref{eq:OM}), which substantially overshadows the second term \cite{supmat}. Fig.~\ref{fig:2}f demonstrates that at higher hole densities ($n_e<-4.2\times 10^{11}$cm$^{-2}$), the orbital magnetization of the majority valley outweighs that of the minority valley, leading to $M>0$. In contrast, at lower hole densities ($-4.2\times 10^{11}$cm$^{-2}<n_e$), the orbital magnetization of the minority valley prevails, resulting in $M<0$. As densities decrease further, $M$ reaches a non-analytic point at $n_e=n_{ALT}$.
%jumps discontinuously.
%and lead to a  derivative $dM/dn_e$. 
%We ascribe this non-analiticity to the behavior of the inverse compressibility $d\mu/dn_e$ at the Lifshitz transition point, see Fig.~2b. This discontinuity in $d\mu/dn_e$ stems from the substantial enhancement of the exchange energy resulting from the creation of a new inner Fermi surface. The relationship between these quantities can be understood by considering the thermodynamic free energy of an itinerant ferromagnet:
% $M_s=\gamma S$ is correct (Mainak) source: Wiki

The non-analytic behavior of $M$ and $\mu$ at $n_{ALT}$ are related, and their  
relationship can be elucidated through the following thermodynamic free energy:
\begin{equation}
    dF = \mu dn_e - (M_s+M )dB.
\end{equation}

The magnetization from spin moments, $M_s$, is simply proportional to the groundstate spin polarization $S$, $M_s=\gamma S$, with $\gamma$ being the gyromagnetic ratio. The valley-Ising metal within the $n_e-U$ phase space under study saturates both the groundstate spin and valley polarization, leading to $\partial M_s/ \partial n_e= \gamma \partial S/ \partial n_e=0$. However, $\partial M/\partial n_e
\neq0$ because there is no such simple relationship exists between the orbital magnetization and the groundstate valley polarization, c.f.~Eq.~\eqref{eq:OM}. In fact, $\partial M/\partial n_e$ is directly proportional to the inverse compressibility $\partial \mu/\partial n_e$ as demonstrated by the following thermodynamic relation:
\begin{equation}
    \frac{ \partial M}{ \partial n_e} = \frac{\partial^2 F}{\partial n_e\partial B}=\frac{\partial M}{\partial \mu}\frac{\partial \mu}{\partial n_e}.
\end{equation} 
%It is noteworthy that the discontinuity of the derivative:
%\begin{equation}
%\frac{d}{dn_e} \left(\frac{M}{n_e}\right)= \frac{1}{n_e} \frac{\partial M}{\partial \mu}\frac{\partial \mu}{\partial n_e} - \frac{M}{n_e^2},
%\end{equation}
%is more pronounced than the discontinuity in $d\mu/dn_e$. This is because the $\frac{1}{n_e} \frac{\partial M}{\partial \mu}$ 
%receives contributions from both $\frac{1}{n_e}\frac{dM}{dn_e}$ and $M/n_e^2$, and their singularity adds up at the ALT.
%%%%%
%
The strong singularity of $ \partial M/ \partial n_e $ close to ALT can be experimentally corroborated by measuring $\mu$ as a function of $B$, utilizing the Maxwell relation $ \partial M/ \partial n_e = - \partial \mu/ \partial B$.

The non-analyticity of $\mu$ and $M$ at the ALT is found to be dependent on the range of repulsive interaction \cite{supmat}. For the dual-gated Coulomb interaction we considered, $\mu$ and $M$ undergo discontinuous jumps at $n_e=n_{ALT}$, a characteristic feature of a first-order phase transition. On the other hand, for ultra-short-range interactions where the repulsive interaction is independent of wavevector, $\mu$ and $M$ are continuous, but $d\mu/dn_e$ and $dM/dn_e$ show a discontinuous leap, a feature of a second-order phase transition.

%\begin{equation}
%    \lim_{n_e\rightarrow n_{ALT}+0^+} dM/d\mu \neq \lim_{n_e\rightarrow n_{ALT}-0^+} dM/d\mu.
%\end{equation}This divergence amplifies the singularity of $\frac{dM}{dn}$ an effect which is further detailed in \cite{supmat}.
%
\begin{figure*}[t]
    \centering
    \includegraphics[width=2\columnwidth]{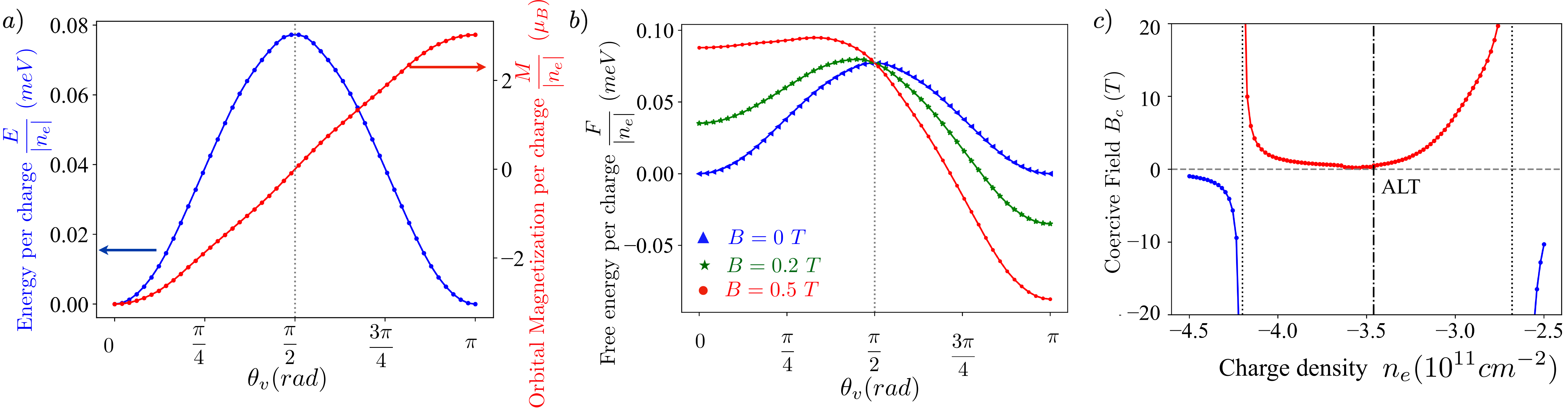}
    \caption{$\vec{a)}$ Energy $E$ and orbital magnetization $M$ per charge carrier vs $\theta_v$ at $(n_e,U)=(-3.6\times 10^{11} cm^{-2},43 ~meV)$. $\theta_v$ is defined in Eq.~\eqref{eq:lagrange}. $\vec{b)}$ The free energy $F=E-MB$ becomes more asymmetrical with increasing magnetic field $B$. $\vec{c)}$ Estimation of coercive field along the horizontal dashed line in Fig. ~\ref{fig:1} and \ref{fig:2} $(U=43~ meV)$. The coercive field diverges when $M=0$ and become a minimum close to the ALT.}
    \label{fig:3}
\end{figure*}

\textit{Collective Valley-rotation and Magnetization Reversal: }We explore the energy and orbital magnetization landscape as a function of global rotation of the valley pseudospin. To generate this rotation, we introduce a Lagrange multiplier term to Eq.~\eqref{eq:H_MF}:
\begin{equation} \label{eq:lagrange}
    \hat{H}_L(\lambda,\theta_v) = \lambda (\cos(\theta_v)\tau_z+\sin(\theta_v)\tau_x),
\end{equation}
where $\lambda$ represents the strength of the coupling and $0\leq \theta_v \leq \pi$ is the polar angle of the valley Bloch sphere. We converge a density-matrix with respect to $\hat{H}+\hat{H}_L $ and compute the total energy per area $E(\theta_v)=\int \frac{d^2k}{(2\pi)^2}\text{Tr}\big[ \hat{\rho}_{\vec{k}}(\theta_v)(\hat{T}_{\vec{k}}+\frac{1}{2}\hat{\Sigma}^F_{\vec{k}})\big]$ and orbital magnetization using Eq.~\eqref{eq:OM}.
%The orbital magnetization only depends weakly on $\lambda$.
% and can be extrapolate to $\lambda=0$ \cite{supmat}.  
In our analysis, we focus on the case where the azimuthal angle $\phi_v=0$, as the energy is degenerate due to the commutation relation $[H,e^{i\phi_v\tau_z}]=0$, see Ref.~\cite{huang2021current} for more discussion.

Figure \ref{fig:3}a~illustrates the behavior of $E(\theta_v)$ and $M(\theta_v)$ v.s. $\theta_v$ at $(n_e,U)=(-3.6\times 10^{11} cm^{-2},43~ meV)$. Notably, all states, except for $\theta=\pi/2$, break orbital time reversal symmetry, while states other than $\theta=0,\pi$ exhibit a tripling of the unit cell area to enable the mixing of states from opposite valleys. We emphasized that the energy required to reverse the valley pseudospin of a single occupied state is on the order of the exchange energy, approximately $20$ meV (as shown in Fig. \ref{fig:2}c,d). In contrast, the energy cost per hole associated with the collective rotation of all valley pseudospins is approximately two orders of magnitude smaller, around $0.1$ meV (c.f. Fig.~\ref{fig:3}a.). This energy disparity bears resemblance to itinerant electron magnets like Ni where exchange energy is $\approx 300$meV ~ and magnetic aniostropic energy is $\approx 3 \mu$eV
\cite{fritsche1987relativistic}.

Next, we generalize the Stoner-Wohlfarth model to study orbital magnetization reversal. Considering a specific point in the $n_e-U$ space, we analyze the following free energy per area
\begin{equation}
    F(\theta_v) = E(\theta_v) - M(\theta_v)B,
\end{equation}
which consists of the even internal energy $E(\theta_v)$ and the odd magnetization energy $M(\theta_v) B$.
As demonstrated in Fig.~\ref{fig:3}b), we found three stationary solutions at $B=0$: two degenerate minima at $\theta_v=0,\pi$ and a maximum at $\theta_v=\pi/2$. With an increase in $B$ in the direction that favors the $\theta_v=\pi$ state, the degeneracy between $\theta_v=0$ and $\theta_v=\pi$ lifts, and the maximum at $\theta_v=\pi/2$ moves towards the higher-energy minimum at $\theta_v=0$. When $B$ attains the coercive field $B_c$, the $\theta_v=0$ state becomes unstable, which prompts a reversal in the ground state order parameter.

The precise theoretical determination of $B_c$ is complex due to its dependency on temperature, the timescale of experimental measurements, and other microscopic details which remain beyond our control. 
Nevertheless, we present a simple estimation of $B_c$ which equates the free energy difference between the two Ising states, $F(\theta_v=\pi)-F(\theta_v=0)= 2M(\theta_v=0)B_c$, to a specified threshold energy $E_{\text{th}}$:
%
%Nevertheless, we present two estimations of $B_c$ and both qualitatively agree with each other. The most straightforward estimation of $B_c$ equates the free energy difference between the two Ising states to a specified threshold energy $E_{\text{th}}$:
% \begin{equation}
%     F(\theta_v=\pi)-F(\theta_v=0)= 2|M(\theta_v=\pi)|B_c  =  E_{\text{th}}.
% \end{equation}
\begin{equation}
2M(\theta_v=0)B_c  =  E_{\text{th}}.
\end{equation}
A logical choice for $E_{\text{th}}$ is $E(\theta_v=0)-E(\theta_v=\pi/2)$, leading to the result shown in Fig.~\ref{fig:3}c.
A unique characteristic of the orbital Chern metal under study is the divergent behavior of $B_c$ towards $\pm \infty$ when $M$ switches sign \cite{polshyn2020electrical}, and its attainment of a minimum when $M$ reaches its peak at the ALT. \footnote{We have not focused on spin moments thus far because they merely yield a uniform magnetization of 1 Bohr magneton per hole (assuming g-factor $g=2$) in the valley-Ising quarter metal.}
% In Ref.~\cite{supmat}, we estimate $B_c$ based on the condition wherein the curvature at $\theta_v=\pi$ alters its sign, ($\frac{\partial^2 F}{\partial \theta^2} \leq 0$), resulting in a qualitatively similar outcome but with larger uncertainties due to numerical differentiation.  

%We have not focused on spin moments thus far because they merely yield a uniform magnetization of 1 Bohr magneton per hole (assuming g-factor $g=2$) in the valley-Ising quarter metal.

%Up to this point, we have not focused on spin moments, as they merely yield a uniform magnetization of 1 Bohr magneton per hole (assuming g-factor $g=2$) in valley-Ising quarter metal. The effect spin-moments and orbital moments are locked by spin-orbit coupling.

\textit{Discussion and outlook:}  
The recently discovered systems of multilayer graphene, perturbed by either the moiré potential or/and dielectric displacement field, demonstrate a unique form of ferromagnetism
\cite{zhou_half_2021,zhou_isospin_2021,seiler2022quantum,han2023correlated,he2021competing,chen2021electrically,tschirhart2021imaging,sinha2022berry,pantaleon2022superconductivity}. In these systems, orbital moments can order through exchange-induced valley polarization,  thereby playing a crucial role in shaping the overall magnetization. Notably, we found that the orbital magnetization $M$ exhibits sign changes and non-analytic behavior, even while the valley polarization remains unchanged. This magnetic behavior starkly deviates from the conventional relationship between spin-magnetization $M_s$ and spin-polarization $S$: $M_s=\gamma S$, where $\gamma$ is the gyromagnetic ratio.
At its core, this difference represents two distinct mechanisms by which a ferromagnetic ground state adjusts its energy in response to an external magnetic field, as per the thermodynamic relation: $dE=-(M + M_s)dB$. The orbital magnetization is associated with the leading order variations in the quasiparticle wavefunctions, $\ket{\delta \psi_{n\vec{k}}}$, induced by the (orbital) magnetic field. Conversely, spin magnetization corresponds directly to the ground state's spin-polarization, which is not associated with changes in the quasiparticle wavefunction or quasiparticle energy. Unlike spin-polarization is a property of the occupied states only, $\ket{\delta \psi_{n\vec{k}}}$ depends on properties of the entire eigenspectrum, including both occupied and unoccupied states. As such, as the valley-Ising quarter metal traverses the $n_e-U$ parameter space, adjustments to the quasiparticle wavefunctions — such as those required to increase layer polarization \cite{supmat} — without modifying the spin-polarization can prompt a significant change in orbital magnetization, all while leaving the spin magnetization undisturbed.

Our prediction regarding the sign-changing behavior of $M$ can be verified by observing the divergence of the coercive field, while the divergent characteristic of $\partial M/\partial n_e$ at the ALT can be corroborated by measuring $\partial \mu/\partial B$ and Shubnikov-de Haas oscillations.

The softness of the collective valley rotation, coupled with the interesting landscape of $M$ in the $n_e-U$ space,indicates that the microscopic dynamics of metallic orbital magnetization warrants a deeper investigation. Our work paves the way for applications in novel electrical control of metallic orbital magnetization, offering a compelling outlook for the creation of future (valleytronic) devices.
%Although our focus is on the quarter metal phase of ABC trilayer graphene, the physics of orbital magnetization discussed here is a characteristic of any metal resulting from doping a Chern band.

%\begin{acknowledgments}
\textit{Acknowledgments:}  
We acknowledge useful discussions with Long Ju, Allan MacDonald, Yafei Ren, Nemin Wei and Andrea Young. We are grateful to the University of Kentucky Center for Computational Sciences and Information Technology Services Research Computing for their support and use of the Morgan Compute Cluster and associated research computing resources.
%\section{Supplementary Materials:--}
%Bulk vs Edge contribution of OM (Appendix)

\bibliography{references}

%%%%%%%%%% Merge with supplemental materials %%%%%%%%%%
\newpage
\widetext
\begin{center}
\textbf{\large Supplementary Materials: Unconventional metallic magnetism: non-analyticity and sign changing behavior of Orbital magnetization in ABC trilayer graphene}
\end{center}
%%%%%%%%%% Merge with supplemental materials %%%%%%%%%%
%%%%%%%%%% Prefix a "S" to all equations, figures, tables and reset the counter %%%%%%%%%%
\setcounter{equation}{0}
\setcounter{figure}{0}
\setcounter{table}{0}
\setcounter{page}{1}
\makeatletter
\renewcommand{\theequation}{S\arabic{equation}}
\renewcommand{\thefigure}{S\arabic{figure}}
%\renewcommand{\bibnumfmt}[1]{[S#1]}
%\renewcommand{\citenumfont}[1]{S#1}

%%%%%%%%%%%%%%%%%%%%%%%%%%%%%%%%%%%%%%%%%%%%%%%%%%%%%%%%%%%%%%%%%%%%%%%%%%%%%%%%%%%%%%%%%%
\maketitle

This supplementary material is divided into four sections. In the first section, we follow Ref.~\cite{shi2007quantum} to formulate an expression for the orbital magnetization in terms of Bloch wavefunctions and their energy dispersion.
In the second section, we disregard the exchange self-energy and elaborate on the layer polarization and orbital magnetization of a (non-interacting) valley-projected Hamiltonian as functions of the chemical potential.
In the third section, we expand our discussion of the orbital magnetization in the quarter metal phase, concentrating on both the bulk and edge contributions.
In the final section, we discuss the association between the non-analyticity of the chemical potential at the annular Lifshitz transition and the range of the repulsive interaction.

\section{A: Derivation of Orbital Magnetization from first order perturbation theory
}

The derivation presented here is based on the work by J. Shi et al. \cite{shi2007quantum}. The thermodynamic definition of  magnetization (magnetic moment per unit area) is given by the familiar expression:
\begin{align}
\vec{M} = -\frac{1}{\mathcal{A}}\left(\frac{\partial \Omega}{\partial \vec{B}}\right)_{T,\mu}  
\end{align}
where $\mathcal{A}$ is the area of the sample and $\Omega=E-TS-\mu N=-pV$ is the grand potential. Since the zero-temperature magnetization
\begin{align}
\tilde{\vec{M}} = -\frac{1}{\mathcal{A}}\left(\frac{\partial K}{\partial \vec{B}}\right)_{\mu}   \label{eq:M_def}
\end{align}
where $K=E-\mu N$ is related to the finite temperature magnetization through the relationship,
$
\tilde{\vec{M}} = \frac{\partial(\beta \vec{M})}{\partial\beta} 
$,
we will first compute $\tilde{\vec{M}}$ then infer $\vec{M}$ from this relationship. Here $\beta = \frac{1}{k_B T}$ and $k_B$ is the Boltzmann constant. In the absence of spin-orbit coupling, we can study spin and orbital magnetization separately. From here and what follows, we only focus on orbital magnetization.

To arrive at an expression for $\tilde{M}$ that involves the quantum mechanical wavefunctions and energy dispersion of a particular crystals, it proves useful to represent Eq.~\eqref{eq:M_def} in integral form:
\begin{align}
\delta K = \int d\vec{r} \delta K(\vec{r}) \equiv - \int d\vec{r} \tilde{\vec{M}}(\vec{r})\cdot \boldsymbol{B}(\vec{r}), \label{eq:delta_K}
\end{align}
where $\delta K(\vec{r})$ is the local change of the energy induced by external magnetic field. 
%and $\tilde{\vec{M}}(\vec{r})$ is the local magnetization.

Let us consider the following generic Hamiltonian,
\begin{align}
\hat{H} &= \hat{H}_{0} + \hat{V}_B, \\
\hat{H}_{0}&= \sum_{n\vec{k}} \epsilon_{n\vec{k}} \ket{ \psi_{n\vec{k}} } \bra{\psi_{n\vec{k}}}, \\
\hat{V}_B &= \frac{e}{2}\left(\hat{\vec{v}}\cdot \vec{A} + \vec{A}\cdot \hat{\vec{v}}\right),
\end{align}
where $\hat{H}_{0}$ is diagonalized by a set of Bloch waves with wavefunction $\psi_{n,\vec{k}}(\vec{r}) \equiv \langle \vec{r} |\psi_{n,\vec{k}}\rangle = e^{i\vec{k}\cdot\vec{r}}u_{n,\vec{k}}(\vec{r})$ and energy dispersion $\epsilon_{n,\vec{k}}$. 
Here $\vec{k}$ are two-dimensional crystal momentum defined in the first Brillouin zone, $\hat{\vec{v}} = \frac{i}{\hbar}\left[\hat{H},\hat{\vec{r}}\right]$ is the velocity operator and $\vec{A}$ is the vector-potential which gives rise to the applied magnetic field $\vec{B}=\vec{\nabla}\times \vec{A}$. 

In the case where the magnetic flux is a rational multiple $p/q$ of the flux quantum per unit cell area, the Hamiltonian $\hat{H}$ can be diagonalized by a collection of magnetic Bloch states within a Brillouin zone that is $q$-times smaller than the Brillouin zone of $H_0$. However, our interest lies primarily in the weak-field response limit, and so we may use first-order (many-body) perturbation theory to estimate the variation in total energy $\delta K$ to a linear order in $V_B$:
\begin{align} \label{eq:delta_K_qm}
\delta K = \sum_{n,\vec{k}} \left[ n_F(\delta 
\epsilon_{n,\vec{k}})\langle\psi_{n,\vec{k}}|\hat{K}_0|\psi_{n,\vec{k}}\rangle + n_F(\epsilon_{n,\vec{k}})\langle\psi_{n,\vec{k}}|\hat{V}_B|\psi_{n,\vec{k}}\rangle+ n_F(\epsilon_{n,\vec{k}})\left(\langle\delta \psi_{n,\vec{k}}|\hat{K}_0|\psi_{n,\vec{k}}\rangle+\langle\psi_{n,\vec{k}}|\hat{K}_0|\delta\psi_{n,\vec{k}}\rangle\right)\right],
\end{align}
%where the leading order change in energy and wavefunctions are
\begin{equation}
   \delta \epsilon_{n\vec{k}} = \langle\psi_{n,\vec{k}}|\hat{V}_B|\psi_{n,\vec{k}} \rangle,
\end{equation}
\begin{align}
|\delta \psi_{n,\vec{k}}\rangle = \sum_{n'\neq n,\vec{k'}}\frac{\langle\psi_{n',\vec{k'}}|\hat{V}_B|\psi_{n,\vec{k}}\rangle}{\left(\epsilon_{n,\vec{k}}-\epsilon_{n',\vec{k'}}\right)} |\psi_{n',\vec{k'}}\rangle \; +
\quad \sum_{ n,\vec{k'}\neq \vec{k}} \frac{\langle\psi_{n,\vec{k'}}|\hat{V}_B|\psi_{n,\vec{k}}\rangle}{\left(\epsilon_{n,\vec{k}}-\epsilon_{n,\vec{k'}}\right)}|\psi_{n,\vec{k'}}\rangle .
\end{align}

% In order to ensure $V_B$ is
% \textcolor{red}{In order to ensure $V_B$ is small everywhere in the material}, 

Following Ref.\cite{shi2007quantum}, we opt for a field configuration where $V_B$ remains small in all places, as opposed to the more commonly used Landau gauge:
\begin{align} \label{eq:B_n_A}
\vec{B}(\vec{r}) = B \cos(qy) \hat{e}_z, \; \; , \;\;
\vec{A}(\vec{r}) = -B \frac{\sin(qy)}{q} \hat{e}_x.
\end{align}
Here the wavevector $q$ is small and the magnitude of the magnetic field $B$ is a constant. Next, we substitute Eq.~\eqref{eq:B_n_A} into Eq.~\eqref{eq:delta_K_qm} and take the limit of $q\rightarrow 0$ and arrive at the following equation:
\begin{align}
\delta K(\vec{r}) &= - \tilde{M}_z \, B \cos(qy)
\end{align}
where $ \tilde{M}_z$ is independent of $q$. Thus
\begin{align} \label{eq:M_micro}
\tilde{M}_z &= -\frac{2}{B\mathcal{A}} \lim_{q\rightarrow 0} \int  \delta K(\vec{r}) \cos(qy)  d\vec{r}.
\end{align}
This equation enables us to express the orbital magnetization in terms of the microscopic spectrum of $H_0$ by employing first-order perturbation theory. In the following discussion, we provide details into the calculation of $\delta K(r)$.

It's essential to highlight that the first two contributions in Eq.~\eqref{eq:delta_K_qm} vanishes. This is because the expectation value of the current operator vanishes for the ground state and on the Fermi surface. Consequently, the orbital magnetization originates exclusively from the variation in the Bloch-state wavefunction. Let us know separate the third term in n Eq.~\eqref{eq:delta_K_qm} into the inter-band and intra-band contributions:
\begin{align}
\delta K^{(inter)}&=\sum_{n,\vec{k}} n_F(\epsilon_{n,\vec{k}})\sum_{n'\neq n,\vec{k'}}\left[ \frac{\langle\psi_{n,\vec{k}}|\hat{K}_0|\psi_{n',\vec{k'}}\rangle \langle\psi_{n',\vec{k'}}|\hat{V}_B|\psi_{n,\vec{k}}\rangle}{\left(\epsilon_{n,\vec{k}}-\epsilon_{n',\vec{k'}}\right)} +c.c\right] \\
\delta K^{(intra)}&=\sum_{n,\vec{k}} n_F(\epsilon_{n,\vec{k}})\sum_{\vec{k'}\neq \vec{k}}\left[ \frac{\langle\psi_{n,\vec{k}}|\hat{K}_0|\psi_{n,\vec{k'}}\rangle \langle\psi_{n,\vec{k'}}|\hat{V}_B|\psi_{n,\vec{k}}\rangle}{\left(\epsilon_{n,\vec{k}}-\epsilon_{n,\vec{k'}}\right)} +c.c\right].
\end{align}
The matrix element of velocity operator $\hat{\vec{v}}$ is given by the following:
\begin{align}
\langle\psi_{n',\vec{k'}}|\hat{\vec{v}}|\psi_{n,\vec{k}}\rangle = \frac{1}{\hbar} \left[\nabla_{\vec{k}}\epsilon_{n,\vec{k}}\delta_{n,n'} \delta_{\vec{k},\vec{k'}} + \left(\epsilon_{n,\vec{k}}-\epsilon_{n',\vec{k'}}\right)\langle u_{n',\vec{k'}}|\nabla_{\vec{k}}u_{n,\vec{k}}\rangle\right].
\end{align} 
The matrix elements of the perturbation $V_B$ is given by the following:

\begin{align}
\langle\psi_{n',\vec{k'}}|\hat{V}_B|\psi_{n,\vec{k}}\rangle 
&=
\frac{e}{2\mathcal{A}}\int d\vec{r} \psi^{*}_{n',\vec{k'}}(\vec{r})\left[\vec{A}(\vec{r})\cdot \hat{\vec{v}}+\hat{\vec{v}}\cdot \vec{A}(\vec{r}) \right]\psi_{n,\vec{k}}(\vec{r})\notag\\
&=-\frac{eB}{4iq \mathcal{A}}\vec{\hat{e}_x}\cdot\left[\int d\vec{r}e^{i(\vec{k}+\vec{q}-\vec{k'})\cdot \vec{r}}u^{*}_{n',\vec{k'}}(\vec{r})\left( \vec{\hat{v}_{k}} + \vec{\hat{v}_{k'}}\right)u_{n,\vec{k}}(\vec{r}) -\left(\vec{q}\rightarrow-\vec{q}\right)\right] \notag\\
&=-\frac{eB}{4iq}\vec{\hat{e}_x}\cdot\Big[\delta_{\vec{k'},\vec{k}+\vec{q}}\langle u_{n',\vec{k}+\vec{q}}|\left(\vec{\hat{v}_{k}} +\vec{\hat{v}_{k+q}}\right)|u_{n,\vec{k}}\rangle -\left(\vec{q}\rightarrow-\vec{q}\right)\Big]
\end{align}
Here $\langle u_{n',\vec{k}+\vec{q}}|\left(\vec{\hat{v}_{k}}+\vec{\hat{v}_{k+q}}\right)|u_{n,\vec{k}}\rangle =\frac{1}{\mathcal{A}_{UC}}\int_{UC} d\vec{x}~ u^{*}_{n',\vec{k}+\vec{q}}(\vec{x}) \left(\vec{\hat{v}_{k}}+\vec{\hat{v}_{k+q}}\right)\ u_{n,\vec{k}}(\vec{x})$.
%\textcolor{red}{WHAT DO YOU MEAN???} 
%It is worth noting that the dot product is ignored for now. We will consider only the x-component of the velocity operator at the end of the calculation.
Using these expression, we arrive at the following equation for $\delta K^{(inter)}\equiv \int \delta K^{(inter)}(\vec{r}) d\vec{r}  $: 
\begin{align}
\delta K^{(inter)}(\vec{r})&=-\frac{eB}{4iq\mathcal{A}} \vec{\hat{e}_x}\cdot \sum_{n,\vec{k}}
n_F(\epsilon_{n,\vec{k}})\sum_{n'\neq n}\left[\psi^{*}_{n,\vec{k}}(\vec{r})\hat{K}_0 \psi_{n',\vec{k}+\vec{q}}(\vec{r})\frac{\langle u_{n',\vec{k}+\vec{q}}|\left(\vec{\hat{v}_{k}}+\vec{\hat{v}_{k+q}}\right)|u_{n,\vec{k}}\rangle}{\left(\epsilon_{n,\vec{k}}-\epsilon_{n',\vec{k}+\vec{q}}\right)}-\left(\vec{q}\rightarrow-\vec{q}\right)\right]+c.c \notag
\\
&=\Bigg[-\frac{eB}{8iq \mathcal{A}} \vec{\hat{e}_x}\cdot \sum_{n,\vec{k}}
n_F(\epsilon_{n,\vec{k}})\sum_{n'\neq n}\left (\epsilon_{n,\vec{k}}+\epsilon_{n',\vec{k}+\vec{q}}-2\mu\right) e^{i\vec{q}\vec{r}} u^{*}_{n,\vec{k}}(\vec{r})u_{n',\vec{k}+\vec{q}}(\vec{r}) \notag\\
&\quad \times \frac{\langle u_{n',\vec{k}+\vec{q}}|\left(\vec{\hat{v}_{k}}+\vec{\hat{v}_{k+q}}\right)|u_{n,\vec{k}}\rangle}{\left(\epsilon_{n,\vec{k}}-\epsilon_{n',\vec{k}+\vec{q}}\right)}-\left(\vec{q}\rightarrow-\vec{q}\right)\Bigg]+c.c 
\end{align}
where we have used the following equation in the second line
\begin{align}
\psi^{*}_{n,\vec{k}}(\vec{r})\hat{K}_0 \psi_{n',\vec{k}+\vec{q}}(\vec{r}) &= \frac{1}{2}\left(\epsilon_{n,\vec{k}}+\epsilon_{n',\vec{k}+\vec{q}}-2\mu\right)e^{i\vec{q}\vec{r}} u^{*}_{n,\vec{k}}(\vec{r})u_{n',\vec{k}+\vec{q}}(\vec{r}) 
\end{align} 

Next we substitute $K^{(inter)}(\vec{r})$  into Eq.\eqref{eq:M_micro}, and performed the integration 
$\int d\vec{r} \cos(\vec{q}\cdot \vec{r}) e^{i\vec{q}\cdot\vec{r}} u^{*}_{n,\vec{k}}(\vec{r})u_{n',\vec{k}+\vec{q}}(\vec{r})$. This lead to the following expression for the inter-band contribution to the zero-temperature magnetization, $\tilde{M}_z^{(inter)}$:
\begin{align}
\tilde{M}^{(inter)}_z &= \lim_{q\rightarrow 0} \Bigg[\frac{e}{8iq \mathcal{A}} \vec{\hat{e}_x}\cdot\sum_{n,\vec{k}}
n_F(\epsilon_{n,\vec{k}})\sum_{n'\neq n}
\quad \langle u_{n,\vec{k}}|u_{n',\vec{k}+\vec{q}}\rangle \frac{\langle u_{n',\vec{k}+\vec{q}}|\left(\vec{\hat{v}_{k}}+\vec{\hat{v}_{k+q}}\right)|u_{n,\vec{k}}\rangle}{\left(\epsilon_{n,\vec{k}}-\epsilon_{n',\vec{k}+\vec{q}}\right)} \left(\epsilon_{n,\vec{k}}+\epsilon_{n,\vec{k}+\vec{q}}-2\mu\right)-\left(\vec{q}\rightarrow-\vec{q}\right)\Bigg]+c.c .
\end{align}
Following the same steps, we can derive the intraband contribution which combine to give us the following total magnetization:

\begin{align}
\tilde{M}_z &= \lim_{q\rightarrow 0}  \Big[\frac{e}{8iq \mathcal{A}}\vec{\hat{e}_x}\cdot \sum_{n,n',\vec{k}}n_F(\epsilon_{n,\vec{k}})\langle u_{n,\vec{k}}|u_{n',\vec{k}+\vec{q}}\rangle \frac{\langle u_{n',\vec{k}+\vec{q}}|\left(\vec{\hat{v}_{k}}+\vec{\hat{v}_{k+q}}\right)|u_{n,\vec{k}}\rangle}{\left(\epsilon_{n,\vec{k}}-\epsilon_{n',\vec{k}+\vec{q}}\right)}
\left (\epsilon_{n,\vec{k}}+\epsilon_{n,\vec{k}+\vec{q}}-2\mu\right)-\left(\vec{q}\rightarrow-\vec{q}\right)\Big]+c.c  \notag\\
&= \lim_{q\rightarrow 0} \frac{e}{4q \mathcal{A}}\vec{\hat{e}_x}\cdot \sum_{n,n',\vec{k}} \operatorname{Im}\Big[ \frac{ \left (\epsilon_{n,\vec{k}}+\epsilon_{n',\vec{k}+\vec{q}}-2\mu\right)
\langle u_{n,\vec{k}}|u_{n',\vec{k}+\vec{q}}\rangle\langle u_{n',\vec{k}+\vec{q}}|\left(\vec{\hat{v}_{k}}+\vec{\hat{v}_{k+q}}\right)|u_{n,\vec{k}}\rangle}{\left(\epsilon_{n,\vec{k}}-\epsilon_{n',\vec{k}+\vec{q}}\right)} \left(n_F(\epsilon_{n,\vec{k}})-n_F(\epsilon_{n',\vec{k}+\vec{q}})\right)\Big] \notag\\
&=\frac{e}{2 \mathcal{A}\hbar}\vec{\hat{e}_x}\cdot\sum_{n,\vec{k}}\operatorname{Im}\Big[n_F(\epsilon_{n,\vec{k}})\langle \frac{\partial u_{n,\vec{k}}}{\partial \vec{k}}|(\epsilon_{n,\vec{k}}+\hat{H}_0(\vec{k})-2\mu)\times |\frac{\partial u_{n,\vec{k}}}{\partial \vec{k}}\rangle
-(\epsilon_{n,\vec{k}}-\mu)n_F'(\epsilon_{n,\vec{k}})\langle \frac{\partial u_{n,\vec{k}}}{\partial \vec{k}}|(\hat{H}_0(\vec{k})-\epsilon_{n,\vec{k}})\times |\frac{\partial u_{n,\vec{k}}}{\partial \vec{k}}\rangle\Big]
\end{align}
Here $n_F'(\epsilon_{n,\vec{k}})$ represents the derivative of the Fermi-Dirac distribution function $n_F(\epsilon_{n,\vec{k}})$ with respect to $\epsilon_{n,\vec{k}}$. At zero temperature, $(\epsilon_{n,\vec{k}}-\mu)n_F'(\epsilon_{n,\vec{k}})=0$, and the intra-band contribution becomes zero.

In the thermodynamic limit, the momentum spacing becomes continuous and we arrive at the following zero-temperature expression for the orbital magnetization used in the maintext:
\begin{align}
    M &= \frac{e}{2 \hbar}\int \frac{d\vec{k}}{(2\pi)^2} \operatorname{Im}\sum_{n}n_F(\epsilon_{n,\vec{k}})\langle \frac{\partial u_{n,\vec{k}}}{\partial \vec{k}}|(\epsilon_{n,\vec{k}}+\hat{H}_0(\vec{k})-2\mu)\times |\frac{\partial u_{n,\vec{k}}}{\partial \vec{k}}\rangle\\
    & = \frac{e}{2\hbar}\int \frac{d\vec{k}}{(2\pi)^2} \sum_{\substack{n}}n_F(\epsilon_{n,\vec{k}}) \operatorname{Im}\sum_{n'\neq n}\frac{\langle u_{n,\vec{k}}|\frac{\partial \hat{H}_0}{\partial \vec{k}}|u_{n',\vec{k}}\rangle\times \langle u_{n',\vec{k}}|\frac{\partial \hat{H}_0}{\partial \vec{k}}|u_{n,\vec{k}}\rangle}{(\epsilon_{n,\vec{k}}-\epsilon_{n',\vec{k}})^2} (\epsilon_{n,\vec{k}}+\epsilon_{n',\vec{k}}-2\mu) 
\end{align}
Note $M$ is invariant under any gauge transformation of the Bloch-waves: $u_{n\vec{k}}\rightarrow e^{i\theta(\vec{k})}u_{n\vec{k}}$. It is convenient to separate $M$ into a bulk contribution ($M_B$) and edge contribution ($M_E$) $M=M_B+M_E$:
\begin{align}
    M_B &=\frac{e}{2\hbar}\int \frac{d\vec{k}}{(2\pi)^2} \sum_{\substack{n}}n_F(\epsilon_{n,\vec{k}}) \operatorname{Im}\sum_{n'\neq n}\frac{\langle u_{n,\vec{k}}|\frac{\partial \hat{H}_0}{\partial \vec{k}}|u_{n',\vec{k}}\rangle\times \langle u_{n',\vec{k}}|\frac{\partial \hat{H}_0}{\partial \vec{k}}|u_{n,\vec{k}}\rangle}{(\epsilon_{n,\vec{k}}-\epsilon_{n',\vec{k}})^2} (\epsilon_{n,\vec{k}}+\epsilon_{n',\vec{k}}),  \\
    M_E &=\frac{e}{2\hbar}\int \frac{d\vec{k}}{(2\pi)^2} \sum_{\substack{n}}n_F(\epsilon_{n,\vec{k}}) \operatorname{Im}\sum_{n'\neq n}\frac{\langle u_{n,\vec{k}}|\frac{\partial \hat{H}_0}{\partial \vec{k}}|u_{n',\vec{k}}\rangle\times \langle u_{n',\vec{k}}|\frac{\partial \hat{H}_0}{\partial \vec{k}}|u_{n,\vec{k}}\rangle}{(\epsilon_{n,\vec{k}}-\epsilon_{n',\vec{k}})^2} (-2\mu)=\frac{\mu e}{\hbar} \int \frac{d\vec{k}}{(2\pi)^2}\sum_{n} n_F(\epsilon_{n,\vec{k}}) \vec{\Omega}_{n,\vec{k}}.
\end{align}
Here $\vec{\Omega}_{n,\vec{k}}$ denotes the Berry curvature. 
They can be determined separately because $M_B$ is independent of the chemical potential $\mu$ when $\mu$ is inside the gap of an insulator while
$M_e= e\mu C_{occ}/h $ is simply proportional to total Chern number $C_{occ}$ of occupied bands. Thus, the orbital magnetization jumps by a factor
$\Delta M=\frac{e~C_{occ}}{h} E_{gap} $ across an energy gap $E_{gap}$.
% Fig.~1 shows the non-interacting bandstructure and how the valley projected orbital magnetization $(M)$ is changing with chemical potential $(\mu)$.

% %%%%%%%%%%%%%%%%%%%%%%%%%%%%%%%%%%%%%%%%%%%%%%%%%%%%%%%%%%%%%%%%%
% \begin{table}[b]
% \caption{\label{tab:graphene_params_ABC}Tight-binding parameters (in eV) for rhombohedral trilayer graphene, see also Refs.~\cite{zhang_band_2010,koshino_trigonal_2009,zhou_half_2021}.}
% \begin{ruledtabular}
% \begin{tabular}{llllllll}
% $\gamma_0$ & $\gamma_1$ & $\gamma_2$ & $\gamma_3$ & $\gamma_4$ & $U$ & $\Delta$ & $\delta$ \\
% $3.160$ & $0.380$ & $-0.015$ & $-0.290$ & $0.141$ & $0.030$ & $-0.0023$ & $-0.0105$ \\
% \end{tabular} 
% \end{ruledtabular} 
% \end{table}
% %%%%%%%%%%%%%%%%%%%%%%%%%%%%%%%%%%%%%%%%%%%%%%%%%%%%%%%%%%%%%%%%%

\begin{figure}[t]
    \centering
    \includegraphics[width=\columnwidth]{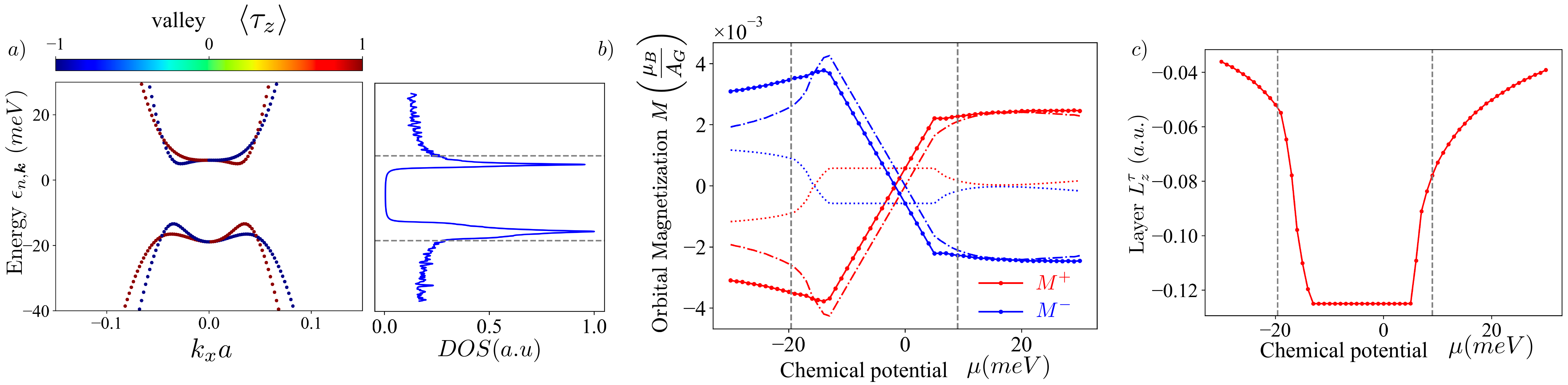}
    \caption{ $\vec{a})$ Bandstructure and density of states (DOS) of valley-projected paramagnetic phase at $U=20~ meV$. Annular Lifshitz transition (ALT) is marked by the dashed gray lines.  $\vec{b})$ Valley-projected orbital magnetization per area of graphene unit cell $(A_G)$ as a function of chemical potential $(\mu)$ at same interlayer potential. $\vec{c})$ Layer polarization $(L_z^\tau)$ v.s. $\mu$ shows that valley-degenerate $L_z$ remains constant inside the insulating gap.}
    \label{fig:rTG_OM_fig4}
\end{figure}

\section{B: Bandstructure, Density of states, Layer polarization and Valley-projected Orbital magnetization of the Paramagnetic state
}

In this section, we discuss the properties of the band Hamiltonian $\hat{T}_{\vec{k}}$ of rhombohedral trilayer graphene (RTG) in the maintext and used the eigenspectrum of $\hat{T}_{\vec{k}}$ to discuss valley-projected orbital magnetization $M^{\tau}$ v.s.~the chemical potential $\mu$. The total magnetization $M^{\tau=+1}+M^{\tau=-1}=0$ because the groundstate of  $\hat{T}_{\vec{k}}$ has time-reversal symmetry.

RTG contains 3 layers of carbon atoms. The interlayer distance is $3.4$~{\AA} and each layer has two sublattices label by $\sigma=A,B$ which is separated by $a=2.46$ {\AA}. The layers of RTG are in a ABC stacking configuration.
Following Refs.~\cite{zhang_band_2010,koshino_trigonal_2009}, we consider a single electronic orbital ($\pi$) per atom, with spin $s=\pm 1/2$, and use the continuum model that can be derived from a tight-binding Hamiltonian by expanding the low-energy dispersion around each valley $\tau$ with $\tau=\pm 1$ at the corners of the Brillouin zone. We label the four flavor combinations of spin $s$ and valley $\tau$ with the flavor index $\alpha\equiv (\tau, s)$. In the basis $\ket{\psi_{\alpha}(\vec{k})}=(\psi_{\alpha A1}(\vec{k}), \psi_{\alpha B1}(\vec{k}), \psi_{\alpha A2}(\vec{k}), \psi_{\alpha B2}(\vec{k}), \psi_{\alpha A3}(\vec{k
}), \psi_{\alpha B3}(\vec{k}))$, the band Hamiltonian is given by
\begin{equation}
\hat{T}_{\vec{k}}=\sum_{s=\pm1,\tau=\pm1} \ket{\psi_{\tau,s}(\vec{k})} h_\tau(\vec{k}) \bra{\psi_{\tau,s}(\vec{k})}.
\end{equation}
The continuum Hamiltonian acting on the orbital space is different in opposite valley but spin-degenerate
\begin{align}\label{eq:continuum_hamiltonian}
    h_\tau(\vec{k}) = \begin{bmatrix}
    t(\vec{k}) + U_1 & t_{12}(\vec{k}) & t_{13} \\
    t_{12}^\dagger(\vec{k}) & t(\vec{k}) + U_2 & t_{12}(\vec{k}) \\
    t_{13}^\dagger & t_{12}^\dagger(\vec{k}) & t(\vec{k}) + U_3
    \end{bmatrix}_{6\times 6}.
\end{align}
Here $\vec{k}=(k_x,k_y)$ is the Bloch momentum measured with respect to the center of valley $\tau$. This Hamiltonian contains matrices with intralayer hopping amplitudes ($t$), nearest-layer hopping amplitudes ($t_{12}$), and next-nearest-layer hopping amplitudes ($t_{13}$), as well as possible potential differences between different layers and/or sublattices due to external gates and/or broken symmetries (terms $U_i$). Explicitly, the various hopping amplitudes are given by
\begin{align}
    t(\vec{k}) = \begin{bmatrix}
      0 & v_0 \pi^\dagger \\
      v_0 \pi & 0
    \end{bmatrix}, \quad 
    t_{12}(\vec{k}) = \begin{bmatrix}
      -v_4 \pi^\dagger & v_3 \pi \\
      \gamma_1 & -v_4 \pi^\dagger
    \end{bmatrix}, \quad
    t_{13} = \begin{bmatrix}
      0 & \gamma_2/2 \\
      0 & 0
    \end{bmatrix}
\end{align}
where $\pi = \tau k_x+i k_y$ is a linear momentum, and $\gamma_i$ ($i=0,\dots, 4$) are hopping amplitudes of the tight-binding model with corresponding velocity parameters $v_i=(\sqrt{3}/2)a\gamma_i/\hbar$. The ${U_i}$ terms are
\begin{align}
    U_1 = \begin{bmatrix}
      \frac{\Delta+\delta+U}{2} & 0 \\
      0 & \frac{\Delta}{2}
    \end{bmatrix}, \quad 
    U_2 = \begin{bmatrix}
      -\Delta & 0 \\
      0 & -\Delta
    \end{bmatrix}, \quad
    U_3 = \begin{bmatrix}
      \frac{\Delta}{2} & 0 \\
      0 & \frac{\Delta+\delta-U}{2}
    \end{bmatrix}
\end{align}

We use the model parameters listed in Table I, which are chosen to match quantum oscillation frequency signatures in Ref.~\cite{zhou_half_2021}.

Next, we diagonalize $h_{\tau}(\vec{k})$ for both $\tau=\pm1$ and discuss the bandstructurte, density of states and the bulk and edge orbital magnetization. 
% The results are shown in Fig.~\ref{fig:rTG_OM_fig4}. 
Fig.~\ref{fig:rTG_OM_fig4}a) shows the valley-projected non-interacting bandstructurte and density of states. The annular Lifshitz transition (ALT) are indicated by dashed gray lines for hole and electron-doping densities. Valley-projected orbital magnetization $(M^{\tau})$ per unit graphene area $(A_G)$ is shown in Fig.~\ref{fig:rTG_OM_fig4}b) as a function of chemical potential $(\mu)$. The dotted lines indicate the bulk contributions $(M^{\tau}_B)$ and the dot-dashed lines indicate the edge contributions $(M^{\tau}_E)$. 
% At the Chern gap, the Edge contribution is directly proportional to the total Chern number of occupied bands as indicted in \textcolor{blue}{Eq.~S.28}.
In Fig.~\ref{fig:rTG_OM_fig4}c), the change in layer polarization $(L_z)$ between the two outer most sublattices $(A_1,B_3)$ as a function of chemical potential indicates the change in wave-functions at the low-energy bands.
% \textcolor{red}{Discuss the results briefly here}

%%%%%%%%%%%%%%%%%%%%%%%%%%%%%%%%%%%%%%%%%%%%%%%%%%%%%%%%%%%%%%%%%
\begin{table}[t]
\caption{\label{tab:graphene_params_ABC}Tight-binding parameters (in eV) for rhombohedral trilayer graphene, see also Refs.~\cite{zhang_band_2010,koshino_trigonal_2009,zhou_half_2021}.}
\begin{ruledtabular}
\begin{tabular}{llllllll}
$\gamma_0$ & $\gamma_1$ & $\gamma_2$ & $\gamma_3$ & $\gamma_4$ & $U$ & $\Delta$ & $\delta$ \\
$3.160$ & $0.380$ & $-0.015$ & $-0.290$ & $0.141$ & $0.030$ & $-0.0023$ & $-0.0105$ \\
\end{tabular} 
\end{ruledtabular} 
\end{table}
%%%%%%%%%%%%%%%%%%%%%%%%%%%%%%%%%%%%%%%%%%%%%%%%%%%%%%%%%%%%%%%%%

\section{C: Layer polarization and different contributions of orbital magnetization in the quarter metal phase
}

In this section, we expand the discussion of orbital magnetization in the quarter metal phase. For the valley-Ising quarter metal, $\theta_v=0,\pi$ in the maintext, we can compute the valley-projected layer-polarization $L_z^{\tau}$, bulk orbital magnetization $M_B^{\tau}$, and edge orbital magnetization $M_E^{\tau}$ separately for both majority valley $\tau=-1$ and minority valley $\tau=+1$. They are given by the following formula:

\begin{align} \label{eq:OM_1}
L_z^\tau &= \sum_{n=1}^{12} \int \frac{d^2k}{(2\pi)^2}
n_F(\epsilon_{n\vec{k}}^{\tau})\bra{\psi_{n\vec{k}}^\tau} \hat{L}_z
\ket{\psi_{n\vec{k}}^\tau}\\
 M^\tau_E =& \; \frac{e\mu}{\hbar}  \sum_{n=1}^{12} \int \frac{d^2k}{(2\pi)^2} 
n_F(\epsilon_{n\vec{k}}^{\tau})\Omega_{n\vec{k}}^{\tau}   \\
M^\tau_B =&  \frac{e }{2 \hbar  } 
 \sum_{\substack{
   n=1}}^{12}
  \int \frac{d^2k}{(2\pi)^2}   \sum_{\substack{n'=1 \\ n' \neq n}}^{12} \bigg[
  n_{F}(\epsilon_{n\vec{k}}^{\tau})(\epsilon_{n,\vec{k}}^{\tau}+\epsilon_{n',\vec{k}}^{\tau}) 
 \frac{ \epsilon_{ij} \operatorname{Im} \left( \langle \psi_{n,\vec{k}}^{\tau}| v_i |\psi_{n',\vec{k}}^{\tau}\rangle  \langle \psi_{n',\vec{k}}^{\tau} |v_j| \psi_{n,\vec{k}}^{\tau} \rangle \right)}{(\epsilon_{n,\vec{k}}^{\tau}-\epsilon_{n',\vec{k}}^{\tau})^2} \bigg].
\end{align}
where $\hat{L}_z=\left(\ket{A_1}\bra{A_1}-\ket{B_3}\bra{B_3}\right)$ measures the polarization between the two outer most sublattice. Note $A_1$ and $B_3$ has no direct (vertical) tunneling  between the nearest layer.

\begin{figure*}[p]
    \centering
    \includegraphics[width=\columnwidth]{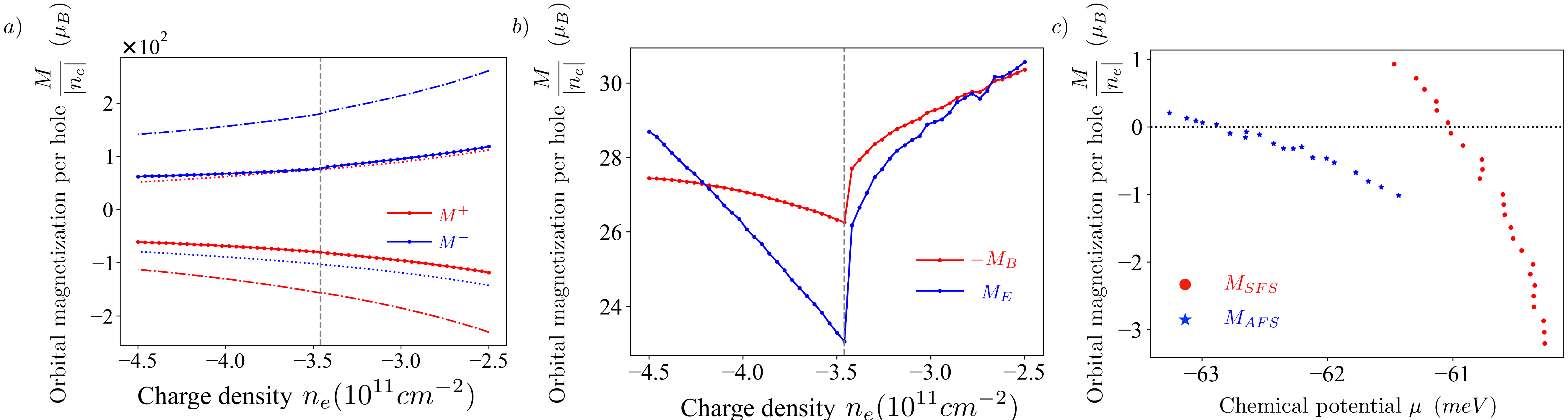}
    \caption{ $\vec{a})$ Valley-projected bulk and edge contributions of orbital magnetization (OM) as a function of density under a constant interlayer potential $(U = 43~\text{meV})$. The dotted line indicates the valley-projected bulk contribution $(M_B^{\tau})$ and the dot-dash line indicates the valley-projected edge contribution $(M_E^{\tau})$. $\vec{b})$ Opposing signs of the total bulk contribution $(M_B = M_B^{+} + M_B^{-})$ and total edge contribution $(M_E = M_E^{+} + M_E^{-})$ persist throughout this regime, with the dominant contribution determining the sign of the total orbital magnetization (OM). $\vec{c})$ Orbital magnetization (OM) exhibits non-analytic behavior at ALT as depicted in its dependence on the chemical potential $(\mu)$, which was determined self-consistently. The red dots correspond to regions in the phase-space $(n_e, U=43~\text{meV})$ where the Fermi surface is simply-connected (SFS), while the blue stars represent regions where the Fermi surface is annular (AFS).}
    \label{fig:rTG_OM_fig5}
\end{figure*}

\begin{figure*}[h]
    \centering
    \includegraphics[width=0.4\columnwidth]{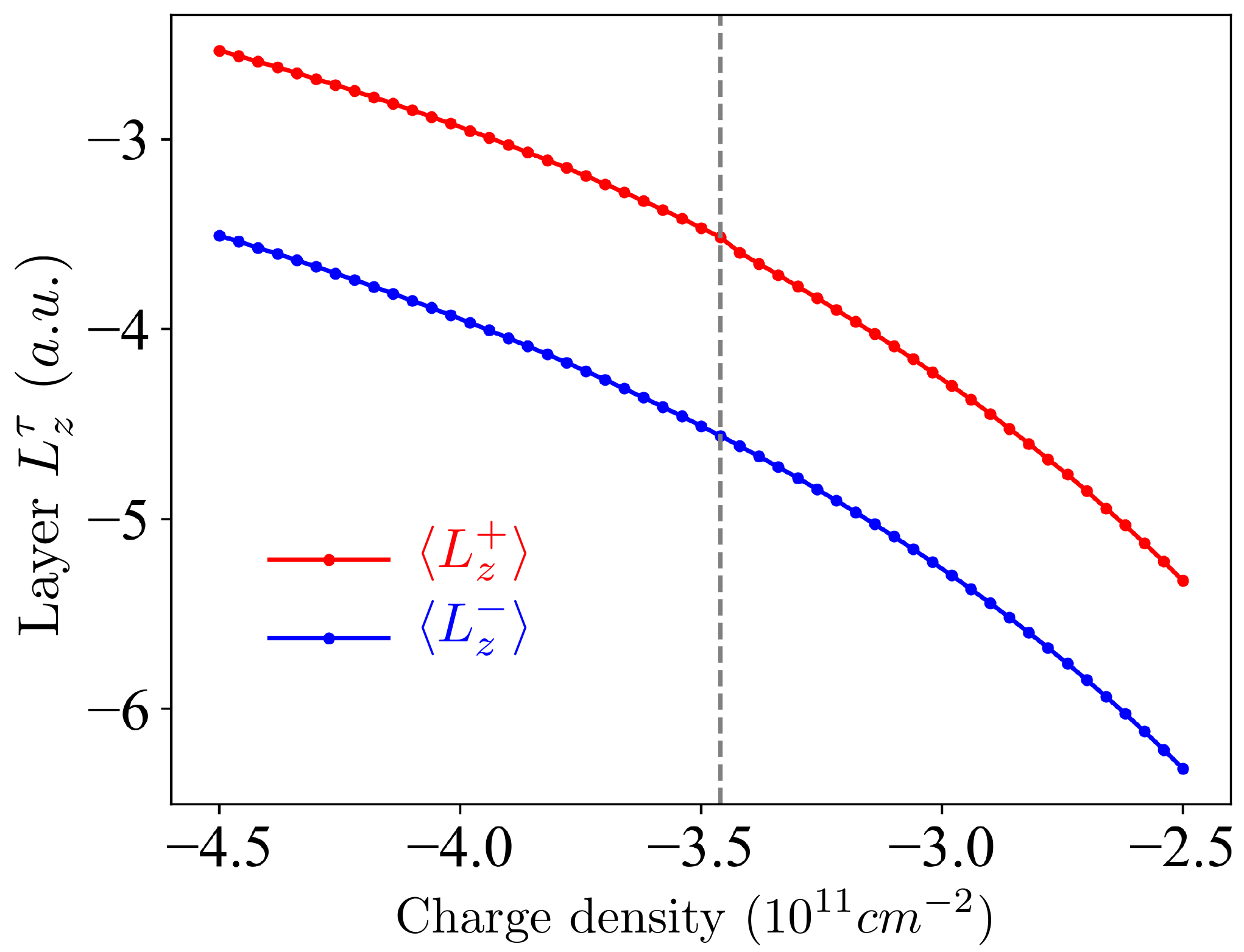}
    \caption{The valley-projected layer polarization $(L_z^{\tau})$ of quarter metal as a function of charge carrier density $(n_e)$. Although valance bands of the majority valley is completely filled throughout this regime, the change in the majority valley layer polarization ($L_z^{-}$) suggests that the wavefunction of majority valley is changing with $n_e$. This observation clarifies the reason behind the change in the majority valley bulk magnetization $(M_B^{-})$ as a function of $n_e$, as illustrated in Fig.~\ref{fig:rTG_OM_fig5}.a.}
    \label{fig:rTG_OM_fig6}
\end{figure*}

Fig.~\ref{fig:rTG_OM_fig5}a) shows $M^\tau_{B}$ and $M^\tau_{E}$ vs $n_e$ in the quarter metal phase. The trend of total magnetization in each valley $M^\tau=M^\tau_{B}+M^\tau_{E}$ as a function of $n_e$ is set by $M^\tau_{E}$. 
Fig.~\ref{fig:rTG_OM_fig5}b) shows the total edge contribution $M_E = M_E^+ + M_E^-$ and the total bulk contribution $M_B = M_B^+ + M_B^-$ v.s. $n_e$. Due to opposite sign, the dominance between total bulk and edge determines the sign of total orbital magnetization. 
Fig.~\ref{fig:rTG_OM_fig5}c) shows the total magnetization for a quarter metal with simply-connected Fermi surface (SFS) and quarter metal with annular Fermi surface (AFS) as a function of chemical potential $(\mu)$. Fig.~\ref{fig:rTG_OM_fig6} shows layer polarization $(L_z^{\tau})$ v.s.~charge carrier density $(n_e)$ in majority and minority valley. At the Ising quarter-metal regime, although the valance band of the majority valley is completely filled and conduction band is empty, the layer polarization of the majority valley $(L_z^{-})$ changes with $n_e$. This suggests that the wavefunction of majority valley is changing with respect to $n_e$, and consequently the majority valley bulk magnetization $(M_B^{-})$ is also changing.

The  abrupt change of orbital magnetization $(M)$ at ALT, as depicted in Fig.~\ref{fig:rTG_OM_fig5}c) is a direct consequence of first order phase transition as we illustrate it in the next section.

\pagebreak 

\section{D: Nature of non-analytic behavior of $\mu$ at the annular Lifshitz transition.
}

In this section, we study the impact of interaction range on the characteristics of the annular Lifshitz transition in the quarter metal phase. Subsequently, we employ a Lagrange multiplier to stabilize excited states, which exhibit a different number of Fermi surfaces in comparison to the ground state.

\begin{figure*}[t]
    \centering
    \includegraphics[width=\columnwidth]{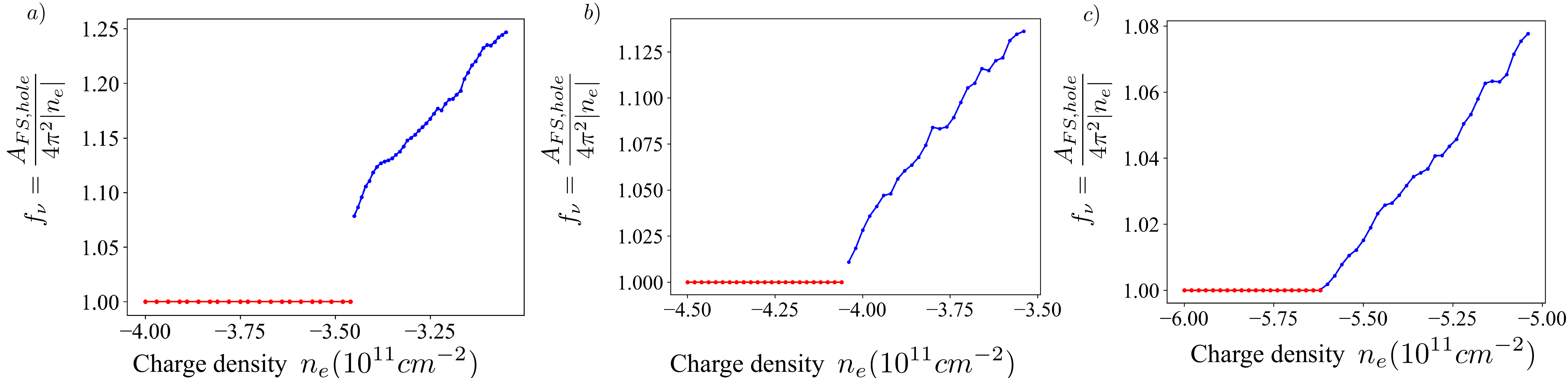}
    \caption{Change in topology of Fermi surface before and after annular Lifshitz transition (ALT) for different ranges of Coulomb interaction: $\vec{a})$ $d=5 nm.$, $\epsilon_r=15$, $\vec{b})$ $d=10 nm.$, $\epsilon_r=30$, $\vec{c})$ Ultra short-range interaction where $V=V_0 \delta(\vec{r})$. This shows a clear pattern that as the interaction range increases, the phase transition happens more abruptly.}
    \label{fig:rTG_OM_fig9}
\end{figure*}

Fig.~\ref{fig:rTG_OM_fig9}a) represents the scenario we explored in the main body of the manuscript. The abrupt jump of $f_\nu$ from the red line (denoting a simple Fermi sea) to the blue line (indicating an annular Fermi sea) signals that the emergence of a new (electron-like) Fermi surface has a minimum critical area.
Fig.~\ref{fig:rTG_OM_fig9}b, c) demonstrates that the nature of this jump is dependent on the specifics of the interaction parameters. 
%As we decrease the range of the Coulomb repulsion (keeping the strength constant) by adjusting the distance between the gates $(d)$ and screening constant $(\epsilon_r)$ shown in Fig.~\ref{fig:rTG_OM_fig9}b), the critical area of the electronic Fermi pocket decreases. 
For ultra-short range Coulomb repulsion where $V=V_0 \delta(\vec{r})$, the change of the Fermi surface area become continuous. 
%continuous change in electronic Fermi pocket area at the ALT suggests a second order phase transition, shown in Fig.~\ref{fig:rTG_OM_fig9}c).
We note that changing $V_0$ will only shift the critical density where ALT happens while the transition remains continuous. 
% 2\pi k_e d/\epsilon_r
% \begin{figure*}[h]
%     \centering
%     \includegraphics[width=0.4\columnwidth]{rTG_OM_fig11.pdf}
%     \caption{Inverse electronic compressibility $\kappa=\frac{\partial \mu}{\partial n_e}$ v.s. charge carrier density $(n_e)$ at $U=43$ meV shows non-analyticity at Annular Lifshitz transition (ALT), marked by the dashed gray line.}
%     \label{fig:rTG_OM_fig11}
% \end{figure*}

To explore the energy landscape of states with simply-connected Fermi surface (SFS) and states with annular Fermi surface (AFS) around ALT, we introduce a Lagrange multiplier $\hat{H}_{l}$ to the mean-field Hamiltonian $\hat{H}$ (Eq.~(1) in the maintext) where
\begin{align}
    \hat{H}_{l}(\vec{k},\lambda)=\lambda e^{-\frac{|\vec{k}|^2}{2\zeta^2}}\sum_{n}|\psi_{n,\vec{k}}\rangle\langle\psi_{n,\vec{k}}|.
\end{align}

\begin{figure*}[h]
    \centering
    \includegraphics[width=0.9\columnwidth]{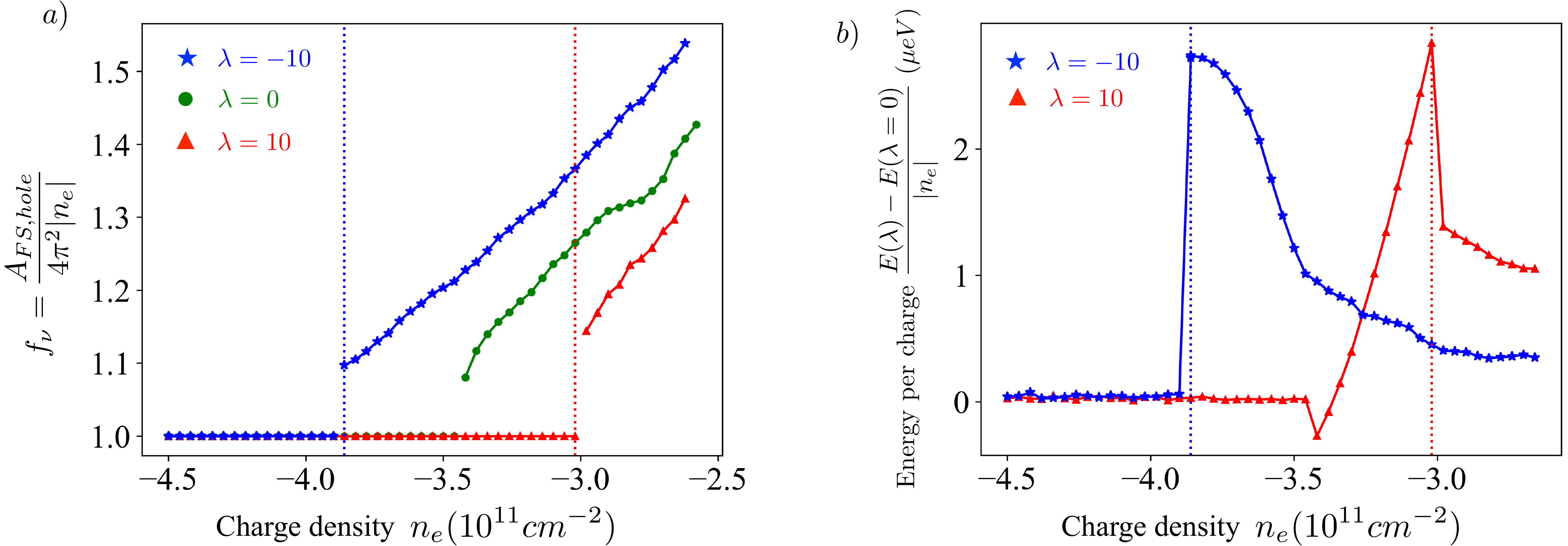}
    \caption{$\vec{a})$ Change in topology of Fermi surface for different Lagrange multiplier: For $\lambda=-10$, the groundstate jumps from SFS to AFS at higher hole density $(n_e<-3.46\times 10^{11}\text{cm}^{-2})$. Conversely, for $\lambda=10$, the groundstate jumps from SFS to AFS at lower hole density $(n_e>-3.46\times 10^{11}\text{cm}^{-2})$. The vertical dotted lines indicate the phase transition at $|\lambda|=10$. $\vec{b})$ Groundstate energy per charge carrier of $\lambda=-10$ and $\lambda=10$ states cross at the $n_{ALT}(|\lambda|)$, and shows a first order phase transition.}
    \label{fig:rTG_OM_fig10}
\end{figure*}

%Here the coupling strength $\lambda>0$ seeds the SFS states where $\lambda=0$ state is AFS and $\lambda<0$ seeds the AFS states where the $\lambda=0$ state is SFS,shown in Fig.~\ref{fig:rTG_OM_fig10}a). 
%
We solve for the eigenspectrum of $\hat{H}+\hat{H}_{l}$ self-consistently for $\lambda=(-10,0,10)$meV at fixed $\zeta=0.1/a$. As shown in Fig.~\ref{fig:rTG_OM_fig10}a), for $\lambda=10$ ($\lambda=-10$), the quarter metal groundstate is a SFS at large hole density and at low density $-3\times 10^{11} \text{cm}^{-2} \lessapprox n_e $ 
($ -3.8\times 10^{11}\text{cm}^{-2} \lessapprox n_e $), it develops a small electron-like Fermi pocket and become an AFS. 
Fig.~\ref{fig:rTG_OM_fig10}b) shows there is an energy crossing at $n_e \approx -3.2\times 10^{11}\text{cm}^{-2}$ where the groundstate jumps from $\lambda=10$ at large hole density to $\lambda=-10$ at small hole density. This confirms the nature of the ALT is a first-order phase transition where the electron-like Fermi pocket appears with a critical area. Note the critical density of ALT depends weakly on  $|\lambda|$.
%groundstate energy of the $\lambda=10$ state is lower (higher) than the $\lambda=-10$ state for $n_e \lessapprox -3.2\times 10^{11}\text{cm}^{-2}$   ($-3.2\times 10^{11}\text{cm}^{-2} \lessapprox n_e$) and there is an energy crossing
%where $n_{ALT}$ couples weakly to the coupling strength $(|\lambda|)$. 

%\bibliography{references}
\end{document}